\documentclass[aps,prapplied,showpacs,twocolumn,preprintnumbers,amsmath,amssymb,longbibliography]{revtex4-2}

\usepackage{graphicx}
\usepackage{dcolumn}
\usepackage{bm}
\usepackage{color}
\usepackage{amsmath}
\usepackage{upgreek}
\usepackage{float}
\usepackage[colorlinks,linkcolor=blue, urlcolor=blue, anchorcolor=blue, citecolor=blue]{hyperref}

\begin{document}

\title{Spatiotemporal differentiators generating optical vortices with transverse orbital angular momentum and detecting sharp change of pulse envelope}

\author{Junyi Huang}
\thanks{These authors contributed equally to this work.}
\affiliation{Interdisciplinary Center of Quantum Information, State Key Laboratory of Modern Optical Instrumentation, and Zhejiang Province Key Laboratory of Quantum Technology and Device, Department of Physics, Zhejiang University, Hangzhou 310027, China}

\author{Jiahao Zhang}
\thanks{These authors contributed equally to this work.}
\affiliation{Interdisciplinary Center of Quantum Information, State Key Laboratory of Modern Optical Instrumentation, and Zhejiang Province Key Laboratory of Quantum Technology and Device, Department of Physics, Zhejiang University, Hangzhou 310027, China}

\author{Tengfeng Zhu}
\affiliation{Interdisciplinary Center of Quantum Information, State Key Laboratory of Modern Optical Instrumentation, and Zhejiang Province Key Laboratory of Quantum Technology and Device, Department of Physics, Zhejiang University, Hangzhou 310027, China}

\author{Zhichao Ruan}
\email{zhichao@zju.edu.cn}
\affiliation{Interdisciplinary Center of Quantum Information, State Key Laboratory of Modern Optical Instrumentation, and Zhejiang Province Key Laboratory of Quantum Technology and Device, Department of Physics, Zhejiang University, Hangzhou 310027, China}

\begin{abstract}
As a new degree of freedom for optical manipulation, recently spatiotemporal optical vortices (STOVs) carrying transverse orbital angular momentums have been experimentally demonstrated with pulse shapers. Here a spatiotemporal differentiator is proposed to generate STOVs with transverse orbital angular momentum. In order to create phase singularity in the spatiotemporal domain, the spatiotemporal differentiator is designed by breaking spatial mirror symmetry. In contrast to pulse shapers, the device proposed here is a simple one-dimensional periodic nanostructure and thus it is much more compact. For a normal incident pulse, the differentiator generates a transmitted STOV pulse with transverse orbital angular momentum. Furthermore, the interference of the generated STOVs can be used to detect the sharp changes of pulse envelopes, in both spatial and temporal dimensions.
\end{abstract}
\maketitle

\section{Introduction}
Optical vortex (OV) is a particular type of optical beams that carries orbital angular momentum (OAM) of photons.\cite{yao2011orbital,franke2008advances,shen2019optical} Typically, by creating phase singularities in two-dimensional transverse electromagnetic fields, the generated OV beams can carry OAM along the longitudinal direction. With such a degree of freedom, longitudinal OAM enables various advanced applications in broadband optical communication,\cite{wang2012terabit,bozinovic2013terabit,lei2015massive} quantum informatics,\cite{ding2015quantum,wang2015quantum,Ming2015Generation} optical tweezers,\cite{paterson2001controlled,padgett2011tweezers} super-resolution imaging,\cite{tamburini2006overcoming,yan2015q} and quantum key distribution\cite{vallone2014free}.

Beyond the spatial domain, recently intense interest has been attracted to explore the OV in the spatiotemporal domain \cite{sukhorukov2005spatio,bliokh2012spatiotemporal, bliokh2015transverse,bliokh2021spatiotemporal}. Remarkably, as an optical pulse, the OAM of spatiotemporal optical vortices (STOVs) can be tilted with respect to the propagation direction and exhibits the transverse components. Recently, such STOVs with transverse OAM have been theoretically proposed and experimentally demonstrated, based on the spatiotemporal control methods  with adjustable resolution and applications.\cite{chong2020generation,hancock2019free,jhajj2017hydrodynamic, jhajj2016spatiotemporal, hancock2021second,wan2020generation, wan2021photonic, wang2021engineering} As a new degree of freedom for optical manipulation, transverse OAM unique in the spatiotemporal domain strongly inspires potential advances to generate STOVs in a novel way.

In this paper, we propose to generate STOVs with an optical spatiotemporal differentiator. Recently, optical analog computation of mathematical differentiation has attracted particular attentions because of its advantages of high throughput and real-time processing.\cite{zangeneh2020analogue} We note that previous optical differentiators separately operate in either the spatial \cite{silva2014performing,bykov2014optical,ruan2015spatial, abdollahramezani2015analog,youssefi2016analog,zhu2017plasmonic,zhang2018implementing,kwon2018nonlocal,guo2018photonic,bykov2018first,cordaro2019high,zhu2019generalized, zhu2020optical, zhou2020flat,huo2020photonic,huang2020two,zhu2021topological} or temporal \cite{liu2008compact,kulishov2005long,rivas2009experimental,bykov2011temporal,kazanskiy2013use,huang2015terahertz,xu2007all,liu2016fully,zhang2014optical,park2007ultrafast} domain. Here we design the differentiator to couple both the spatial and temporal computation in the spatiotemporal domain. Moreover, we show that in order to create phase singularities and generate OVs, it is necessary for the spatiotemporal differentiator to break spatial mirror symmetry. In contrast to the pulse shapers,\cite{chong2020generation,hancock2019free,jhajj2017hydrodynamic} the device we propose here is a simple one-dimensional periodic nanostructure directly implemented to the pulse without spatial or temporal Fourier transform and thus it is much more compact. We demonstrate that for a Gaussian envelope pulse normally incident on the structure, the transmitted pulse is a STOV carrying transverse OAM. Furthermore, the interference of the generated STOVs can be used to detect the sharp changes of pulse envelopes, in both spatial and temporal domains. Our results show the spatial and temporal detection resolutions of 18 $\upmu$m and 182 fs respectively, with great benefits of high-resolution and ultra-fast detection.

\begin{figure}[htbp]
\centerline{\includegraphics[width=3.2in] {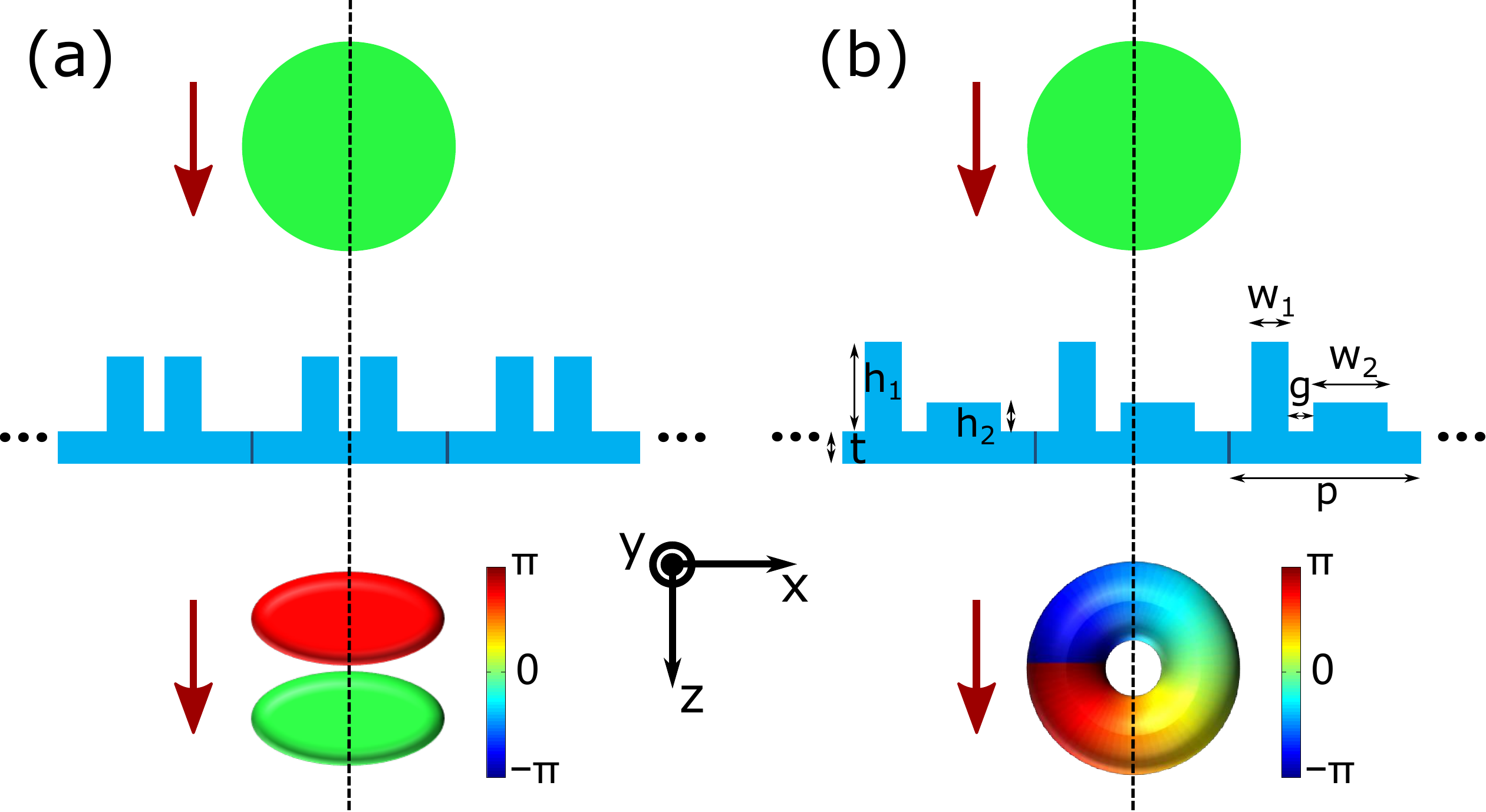}}
\caption{Spatial-symmetry analysis of spatiotemporal differentiator to generate STOVs. (a) If a one-dimensional periodic structure has the mirror symmetry about a plane $x=0$ (indicated by the dashed line), for an incident pulse with both Gaussian envelopes in the spatial and temporal domains, the phase distribution of the transmitted pulse has the same symmetry about the mirror plane, and thus it cannot generate STOVs. (b) Breaking mirror symmetry of a spatiotemporal differentiator is necessary for generating the phase singularity, where the transmitted pulse corresponds to a STOV carrying transverse orbital angular momentum. Moreover, the mirror symmetry cannot exist for any normal incident plane. \label{fig:1}}
\end{figure}

\section{Principle and Design}
Figure \ref{fig:1} schematically shows the spatial-symmetry analysis of spatiotemporal differentiator to generate STOVs. Without loss of generality, we first consider an one-dimensional periodic grating structure with mirror symmetry about the plane $x=0$ (indicated by the dashed line) [Fig.~\ref{fig:1}(a)]. Suppose that a pulse with both Gaussian envelopes in the spatial and temporal domains impinges the structure at normal incidence. Due to the mirror symmetry, the phase distribution of the transmitted pulse must be also symmetric about the mirror plane, and thus there is no phase singularity. Therefore, in order to create a phase singularity and generate STOVs for any incident pulse at normal incidence, it is necessary to break the mirror symmetry of the structure, that is, the mirror symmetry cannot exist for any incident plane. Below we show that by mirror symmetry breaking, a spatiotemporal differentiator can generate STOVs with transverse OAM [Fig.~\ref{fig:1}(b)].

To demonstrate the generation of STOVs, we propose a spatiotemporal differentiator for optical pulses with the center wavelength $\lambda_0=1196$nm, by considering practical experimental demonstration with a femtosecond light source with optical parametric oscillator. Here in order to break the mirror symmetry, we design a spatiotemporal differentiator where the two silicon rods in each period are etched on a silicon substrate with different sizes [Fig.~\ref{fig:1}(b)]. By calculating the transmission spectrum function and inspecting the winding number in the phase diagram, the geometry parameters of the two rods are determined as ${{h}_{1}}=388$ nm, ${{h}_{2}}=160$ nm, ${{w}_{1}}=160$ nm, ${{{w}}_{2}}=432$ nm, and the gap between them is $g=64$ nm. The thickness of substrate and the structure period are $t=20$ nm and $p=1000$ nm, respectively. We note that there are much more degrees of freedom for the parameters to generate STOVs for the normal incident light, by considering the spatial mirror symmetry breaking. Such a device with these suitable geometric parameters can be fabricated through standard lithography and etching processing, with exposing different doses of hydrogen silsesquioxane photoresist.\cite{namatsu1998nano, zhao2021high}

We consider an incident pulse with the $s$-polarization whose electric field is only along the $y$ direction. In order to specifically depict the transformation between the input and the transmitted pulse envelopes, we decompose the incident (transmitted) field into a series of plane waves by Fourier transform as ${{s}_{\text{in}\left( \text{tran} \right)\text{ }}}(x,t)=\iint{{{{\tilde{s}}}_{\text{in}\left( \text{tran} \right)\text{ }}}}\left( {{k}_{x}},\Omega\right)\exp \left( i{{k}_{x}}x-i\Omega t \right)\text{d}{{k}_{x}}\text{d}\Omega$, where ${{s}_{\text{in}\left( \text{tran} \right)}}$ and ${{\tilde{s}}_{\text{in}\left( \text{tran} \right)\text{ }}}$ are the envelope amplitude and the corresponding envelope spectrum of the incident (transmitted) field, respectively. ${{k}_{x}}$ is the wavevector component parallel to the structure interface. $\Omega $ is the sideband angular frequency from the center one ${{\omega }_{0}}$, i.e. $\Omega =\omega -{{\omega }_{0}}$. Then, the pulse envelope transformation is determined by the transmission spectrum function $H({{k}_{x}},\Omega )\equiv {{\tilde{s}}_{\text{tran}}}({{k}_{x}},\Omega)/{{\tilde{s}}_{\text{in}}}({{k}_{x}},\Omega )$.

\begin{figure}[t]
\centerline{\includegraphics[width=3.2in] {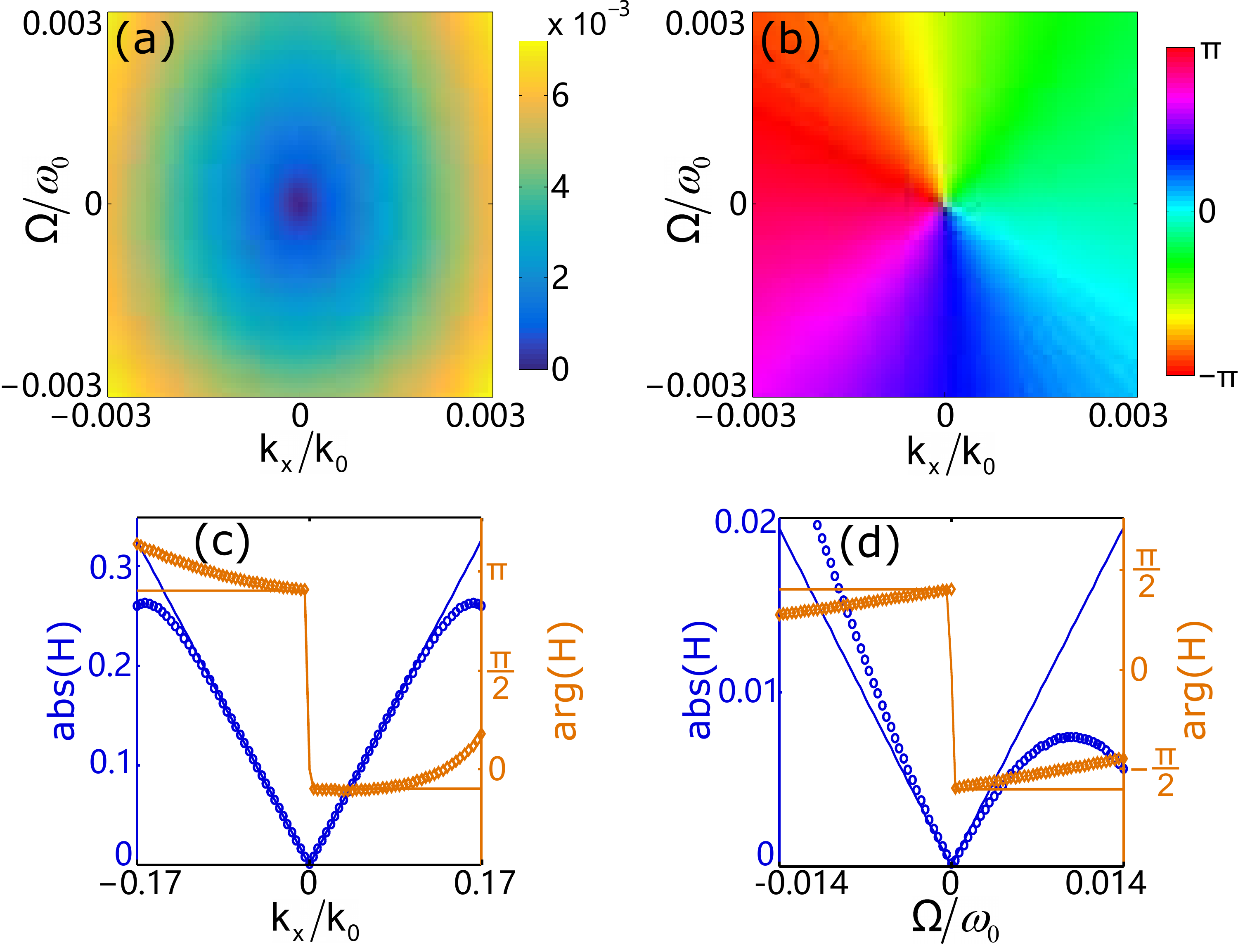}}
\caption{Transmission spectrum function of the designed spatiotemporal differentiator. (a) Amplitude and (b) phase distributions of the spectrum function with respect to $k_x$ and $\Omega$. (c-d) The amplitudes and phases of the spectrum function along (c) $\Omega=0$ and (d) $k_x=0$. The blue circles and lines correspond to the amplitudes of simulation results and the fitting ones with Eq.~(\ref{eq:1}), and the orange rhombi and lines correspond to the phases, respectively. $\omega_0$ and $k_0$ are the frequency and the wavenumber of light in vacuum at the center wavelength $\lambda_0$. \label{fig:2}}
\end{figure}

With the periodicity, our STOV device can only diffract normal incident light in the zeroth order due to ${{2\pi } \mathord{\left/{\vphantom {{2\pi } p}} \right.\kern-\nulldelimiterspace} p} > {k_0}$, where ${k_0}$ is the wavenumber of light in vacuum at the center wavelength ${\lambda _0}$. We numerically calculate the transmission spectrum function $H$ by the finite-element method using the commercial software package COMSOL with frequency domain module and check the convergence of results. In the simulation, the optical constants for silicon are referred to the experimental data in Ref.~\cite{green2008self}. Figures \ref{fig:2}(a) and \ref{fig:2}(b) show the amplitude and phase distributions of the transmission spectrum function with respect to ${{k}_{x}}$ and $\Omega $, respectively. Furthermore, the blue circles and orange rhombi in Figs.~\ref{fig:2}(c) and \ref{fig:2}(d) correspond to the amplitudes and phases of $H$ along $\Omega =0$ and ${{k}_{x}}=0$ respectively. They show that $H$ exhibits good linear dependence on ${{k}_{x}}$ and $\Omega $ about within ${{ - 0.07 \le {k_x}} \mathord{\left/{\vphantom {{ - 0.07 \le {k_x}} {{k_0}}}} \right. \kern-\nulldelimiterspace} {{k_0}}} \le 0.07$ and ${{ - 0.007 \le \Omega } \mathord{\left/{\vphantom {{ - 0.007 \le \Omega } {{\omega _0}}}} \right.\kern-\nulldelimiterspace} {{\omega _0}}} \le 0.007$, and the phase shifts with $\pi$ at the minima occurring at ${{k}_{x}}=0$ and $\Omega =0$, respectively. It indicates that the structure enables the first-order differentiation in both the spatial and temporal domains. Around ${{k}_{x}}=0$ and $\Omega =0$, the transmission spectrum function $H$ has the form
\begin{equation}
H={{C}_{x}}{{k}_{x}}+{{C}_{t}}\Omega,\label{eq:1}
\end{equation}
where ${{C}_{x}}$ and ${{C}_{t}}$ are two complex numbers. We note that the parameter ${{C}_{x}}$ has a phase shift from ${{C}_{t}}$, which leads to the phase singularity in the spectrum domain shown as Fig.~\ref{fig:2}(b).

The blue and orange lines in Figs.~\ref{fig:2}(c) and \ref{fig:2}(d) respectively show the fitting results of amplitudes and phases with Eq.~(\ref{eq:1}), where ${C_x} = 1.92\exp (-0.31i)/{k_0}$ and ${C_t} = 1.39\exp (-1.88i) /{\omega _0}$. Here $\omega_0$ is the frequency at the center wavelength $\lambda_0$. We note that since ${{C}_{x}}/{{C}_{t}}=1.38c \cdot {\exp(1.57i)}$, where $c$ is the velocity of light in vacuum, the winding number of the phase singularity is equal to 1, which means that there must be a zero amplitude in an enclosed loop around ${{k}_{x}}=0$ and $\Omega=0$ shown as Fig.~\ref{fig:2}(a). Importantly, such a phase singularity leads to the asymmetry phase modulations for the plane waves with ${{k}_{x}}<0$ and ${{k}_{x}}>0$ with the same $\Omega$, which can contribute to generating STOVs in the spatiotemporal domain.

We note that for the input signal with ${E_{in}}(x,t) = {s_{in}}(x,t){e^{ - i{\omega _0}t}}$, after propagating through the spatiotemporal differentiator, the output signal is expressed as
\begin{eqnarray}
{E_{tran}}(x,t) & = & {s_{tran}}(x,t){e^{ - i{\omega _0}t}} \nonumber \\
& = & \left( - i{C_x}\frac{\partial {s_{in}}}{{\partial x}} + i{C_t} \frac{\partial {s_{in}}}{{\partial t}} \right) {e^{ - i{\omega _0}t}}.  \label{eq:2}
\end{eqnarray}
Here the output envelope ${s_{tran}}$ is the first order differentiation of the input envelope ${s_{in}}$ in both spatial and temporal domains. Therefore, the STOV generator can be applied to detect sharp changes of pulse envelopes, which is similar to the image processing of edge detection by the spatial differentiators in the real space.\cite{zhu2017plasmonic,zhu2019generalized,zhu2020optical,zhu2021topological}

\section{Generation of spatiotemporal optical vortices}

\begin{figure}[htbp]
\centerline{\includegraphics[width=3.2in] {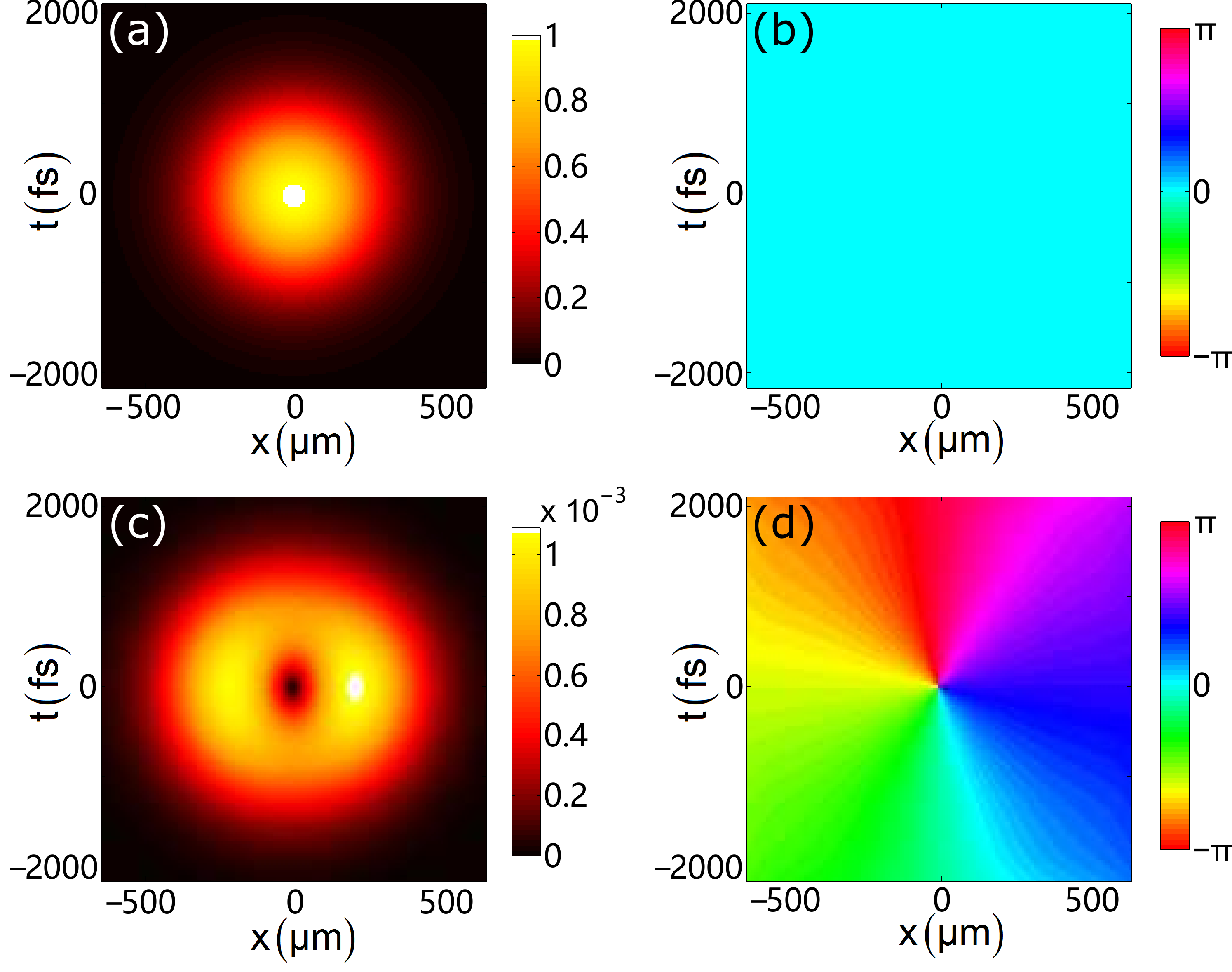}}
\caption{Generation of a STOV by the spatiotemporal differentiator for a Gaussian-enveloped incident pulse. (a) Amplitude and (b) phase distributions of the incident pulse envelope. (c) Amplitude and (d) phase distributions of the transmitted one. The transmitted pulse with transverse OAM has a phase singularity, leading to the zero amplitude at the pulse center.\label{fig:3}}
\end{figure}

To demonstrate the generation of STOVs with transverse OAM, we first simulate an incident pulse with Gaussian envelopes in both spatial and temporal domains. Figure \ref{fig:3}(a) shows the field amplitude distribution of the incident pulse envelope in $(x, t)$ coordinate, where the beam waist and the pulse width are 586 $\upmu$m and 2046 fs respectively, ensuring a spectrum bandwidth within the range shown in Fig.~\ref{fig:2}. The incident pulse has a constant phase distribution shown as Fig.~\ref{fig:3}(b). After passing through the structure, we simulate the transmitted pulse by the Fourier transform method, with the transmission spectrum function in Fig.~\ref{fig:2}. The amplitude and phase distributions of the transmitted pulse envelope are depicted in Figs.~\ref{fig:3}(c) and \ref{fig:3}(d), respectively.

\begin{figure*}[htbp]
\centerline{\includegraphics[width=7.0in] {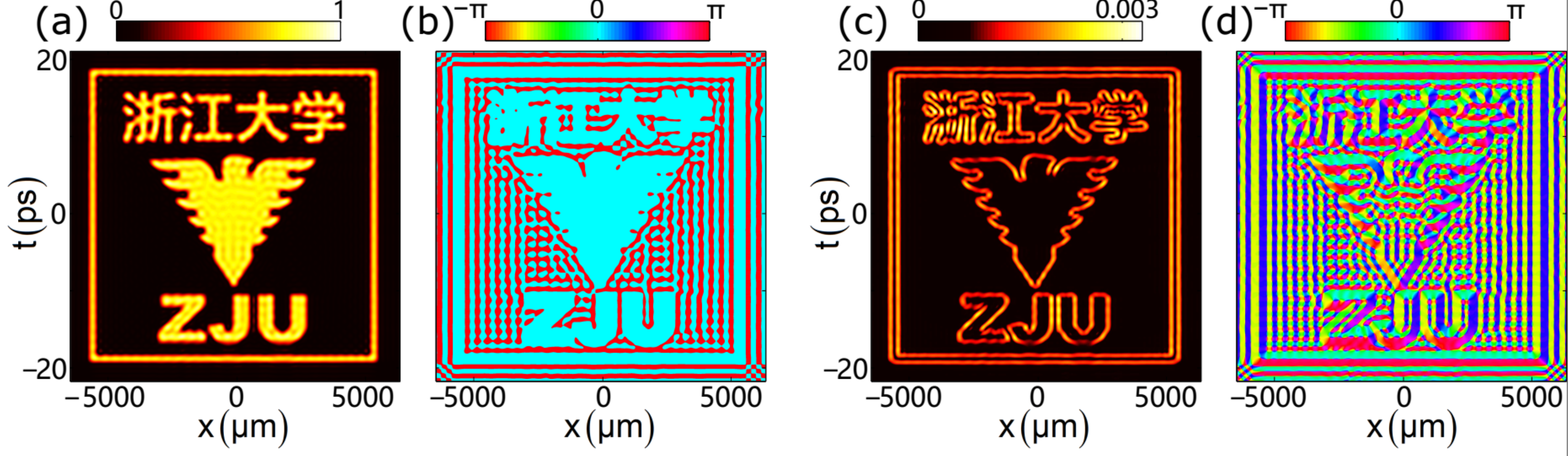}}
\caption{Generation of STOVs for an arbitrary amplitude modulated spatiotemporal pulse. (a) Amplitude and (b) phase distributions of an incident pulse envelope as the Zhejiang University logo. The phases of the incident pulse envelope are binary with only 0 or $\pi$, without phase singularities. (c) Amplitude and (d) phase distributions of the transmitted pulse envelope. \label{fig:4}}
\end{figure*}

We note that the transmitted pulse has a phase singularity at the pulse center [Fig.~\ref{fig:3}(d)], where a STOV exhibits the zero amplitude of the pulse [Fig.~\ref{fig:3}(c)]. In comparison with the symmetrical phase distribution of the incident pulse [Fig.~\ref{fig:3}(b)], the asymmetrical phase of the transmitted pulse envelope originates from the mirror symmetry breaking of the structure. Moreover, as expected, the field amplitude profile of the transmitted pulse in Fig.~\ref{fig:3}(c) indeed shows a typical first-order Hermite-Gaussian profile along each direction, which has a zero value amplitude at the pulse center. Importantly, the vortex phase distribution in Fig.~\ref{fig:3} (d) carries a singularity, corresponding to the central zero amplitude in Fig.~\ref{fig:3} (c). Since the optical vortex exists only in $(x, t)$ coordinate, it carries a transverse orbital angular momentum.

The spatiotemporal differentiator can generate STOVs for arbitrary amplitude modulated spatiotemporal pulses, because the phase singularity in the transmission spectrum function is non-local, akin to the topological spatial differentiator.\cite{zhu2021topological} To further illustrate the STOV generations, here we consider the incident pulse with the amplitude modulation as the Zhejiang University logo in the spatiotemporal domain [Fig.~\ref{fig:4}(a)]. Correspondingly, the phases of the incident pulse envelope are binary with only 0 or $\pi$ [Fig.~\ref{fig:4}(b)], and thus without phase singularities.
We simulate the pulse transmitted through the spatiotemporal differentiator. Figures \ref{fig:4}(c) and \ref{fig:4}(d) correspond to the amplitude and phase distributions of the transmitted pulse envelope, respectively. Remarkably, Fig.~\ref{fig:4}(d) shows the generation of large numbers of adjacent STOVs in the spatiotemporal domain.

Moreover, as the generated STOVs interfere with each other, the amplitude distribution of Fig.~\ref{fig:4}(c) shows that the constructive interference occurs at the sharp changes of the incident pulse envelope in both the spatial and temporal domains, while the destructive one strongly takes place where the amplitudes have slight variations. As shown in Eq.~(\ref{eq:2}), this highlighting-sharpness effect is contributed by the differentiation computing of the STOV generator in the spatiotemporal domain.

\section{Resolution of detecting sharp change}

\begin{figure}[htbp]
\centerline{\includegraphics[width=3.2in] {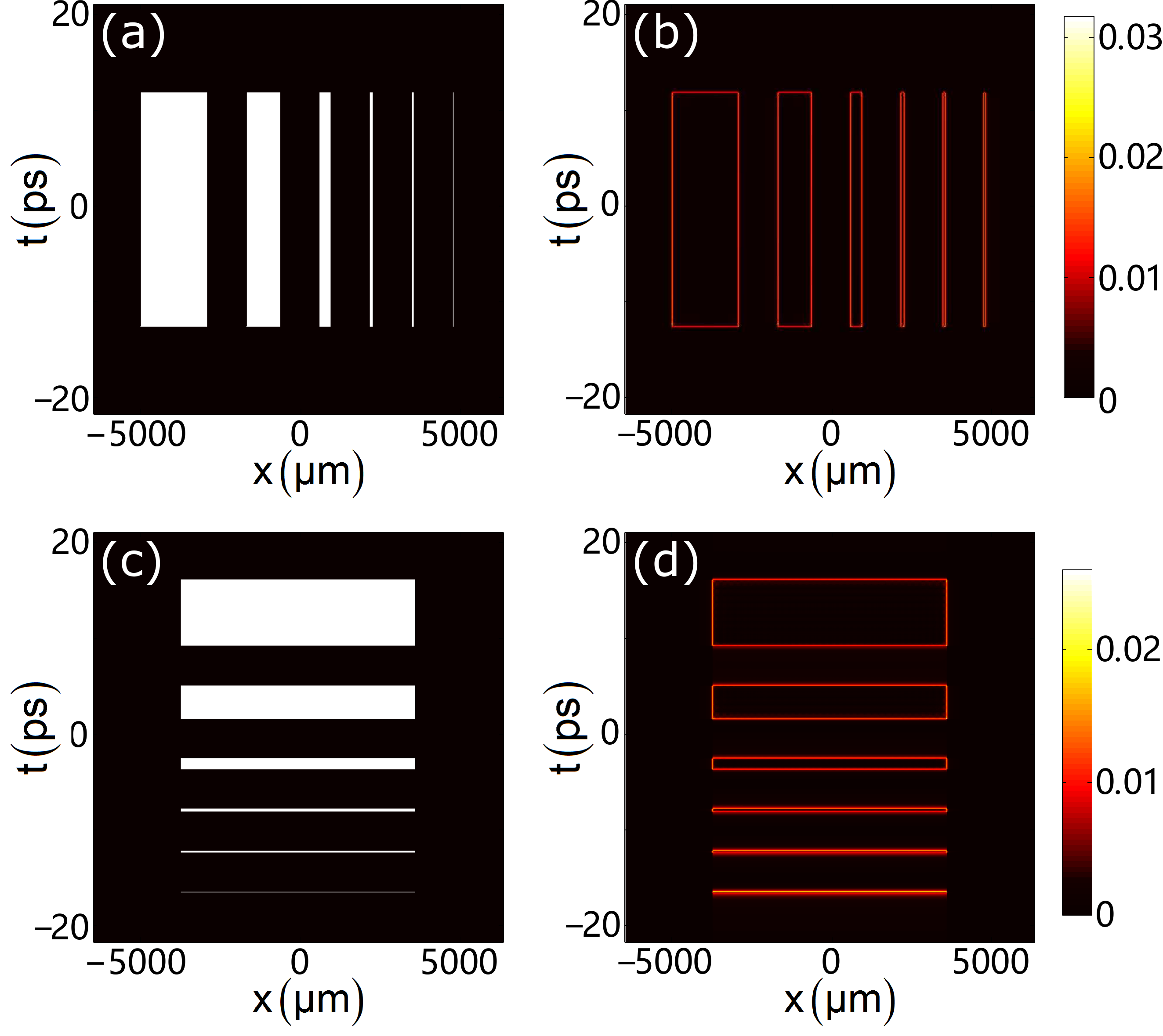}}
\caption{Estimation of spatial and temporal detection resolutions of the spatiotemporal differentiator. (a) The incident rectangular pulses with different spatial widths of 2058, 1038, 346, 91, 55 and 18 $\upmu$m, respectively and the same duration of 24.305 ps. (b) The transmitted pulse corresponding to (a). (c) The incident rectangular pulses with different durations as 6866, 3464, 1155, 304, 182 and 61 fs, respectively and the same spatial profile width of 7286 $\upmu$m. (d) The transmitted pulse corresponding to (c). \label{fig:5}}
\end{figure}

Since the differentiation computing of the STOV generator works within narrow spatiotemporal bandwidths of the transmission spectrum function, we investigate the resolution of detecting sharp changes. We first estimate the spatial resolution by considering rectangular enveloped pulses with a relatively long duration in order to ensure the temporal spectrum within the bandwidth of the STOV generator. Figure \ref{fig:5}(a) shows the incident pulsed fields with different widths of 2058, 1038, 346, 91, 55 and 18 $\upmu$m, respectively, while the signal durations are fixed to 24.305 ps. Figure \ref{fig:5}(b) depicts the amplitude distributions of the transmitted pulses. Indeed, the sharp changes of the pulses are clearly highlighted in both the spatial and temporal domains. However, the left and right spatial edges become blurred, and it is difficult to be separated when the envelope width reduces. From the narrowest width, where the two edges can be distinguished, the spatial resolution of the STOV generator is about 18 $\upmu$m.

Figures \ref{fig:5}(c) and \ref{fig:5}(d) show the estimation of the temporal detection resolution for the STOV generator. The incident pulses are also with rectangular envelopes in both spatial and temporal domains [Fig.~\ref{fig:5}(c)]. In order to estimate the temporal resolution, the incident pulses have different durations as 6866, 3464, 1155, 304, 182 and 61 fs. Meanwhile they have a relatively large width of 7286 $\upmu$m so as to reduce the impact of the spatial bandwidth. Figure \ref{fig:5}(d) exhibits the amplitude distributions of transmitted pulses. We note that the edges of the four long durations in the spatiotemporal domain are clearly detected. In contrast, it is difficult to distinguish the short duration and thus the temporal detection resolution of the STOV generator is about 182 fs.

\section{Conclusion and Discussion}
In summary, we have proposed a spatiotemporal differentiator, which generates STOVs with transverse orbital angular momentum and can be used to detect sharp changes in pulse envelopes. The spatiotemporal differentiator is designed by breaking the mirror symmetry, and the nonlocal effect \cite{kwon2018nonlocal} enables the generations of STOVs for arbitrary modulated spatiotemporal pulses. We note that breaking the mirror symmetry of the device is necessary for the normal incident light, but not sufficient to generate STOVs regarding the winding phase. Furthermore, we show that the interference of the generated STOVs can be used to detect the sharp changes of pulse envelopes, for both spatial and temporal dimensions, with great benefits of high-resolution and ultra-fast optical detection. Practically, the input spatiotemporal modulated pulse could be generated by ultrafast moving objects, and a long duration is preferred in order to alleviate the spatiotemporal diffraction effect.

Also we can generate an opposite topological charge of the phase singularity, by flipping the structure about the plane $x=0$. Moreover, by breaking the mirror symmetry, transverse OAM with high-order topological charges can be generated based on the high-order spatiotemporal differentiation. We note that the designed spatiotemporal differentiator is much more compact than the pulse shapers to generate STOVs. Therefore, it paves the way forward generating and modulating the transverse OAM with integrated devices, which could be important in the applications of STOVs.

\section*{Acknowledgement}
The authors acknowledge funding through the National Key Research and Development Program of China (Grant No. 2017YFA0205700), the National Natural Science Foundation of China (NSFC Grants Nos. 12174340 and 91850108), the Open Foundation of the State Key Laboratory of Modern Optical Instrumentation, and the Open Research Program of Key Laboratory of 3D Micro/Nano Fabrication and Characterization of Zhejiang Province.


%

\end{document}